\documentclass[aps,prl,superscriptaddress,twocolumn,amsmath,amssymb]{revtex4}

\usepackage{color}
\usepackage{graphicx}
\usepackage{bm}
\definecolor{Blue}{rgb}{0.00, 0.00, 1.00}
\definecolor{Red}{rgb}{1.00, 0.00, 0.00}

\begin{document}

\title{Plasmons and screening in monolayer and multilayer black phosphorus}

\author{Tony Low}
\email{tonyaslow@gmail.com}
\affiliation{IBM T.J. Watson Research Center, 1101 Kitchawan Rd., Yorktown Heights, NY 10598, USA}
\affiliation{Department of Electrical Engineering, Yale University, New Haven, Connecticut 06511 }
\author{Rafael Rold\'an}
\affiliation{Instituto de Ciencia de Materiales de Madrid. CSIC. Sor Juana In\a'es de la Cruz 3. 28049 Madrid, Spain}
\author{Han Wang}
\affiliation{IBM T.J. Watson Research Center, 1101 Kitchawan Rd., Yorktown Heights, NY 10598, USA}
\author{Fengnian Xia}
\affiliation{Department of Electrical Engineering, Yale University, New Haven, Connecticut 06511 }
\author{Phaedon Avouris}
\affiliation{IBM T.J. Watson Research Center, 1101 Kitchawan Rd., Yorktown Heights, NY 10598, USA}
\author{Luis Mart\a'in Moreno}
\affiliation{Instituto de Ciencia de Materiales de Arag\a'on and Departamento de F\a'isica de la Materia Condensada,
CSIC-Universidad de Zaragoza, E-50009 Zaragoza, Spain }
\author{Francisco Guinea}
\affiliation{Instituto de Ciencia de Materiales de Madrid. CSIC. Sor Juana In\a'es de la Cruz 3. 28049 Madrid, Spain}

\date{\today}
\begin{abstract}
Black phosphorus exhibits a high degree of band anisotropy. However, we found that its in-plane static screening remains relatively isotropic for momenta relevant to elastic long-range scattering processes. On the other hand, the collective electronic excitations in the system exhibit a strong anisotropy. Band non-parabolicity leads to a  plasmon frequency which scales as $n^{\beta}$, where $n$ is the carrier concentration, and $\beta<\tfrac{1}{2}$. Screening and charge distribution in the out-of-plane direction are also studied using a non-linear Thomas-Fermi model.
\end{abstract}
\maketitle

\emph{Introduction---} Black phosphorus (BP) is one of the thermodynamically more stable phases of phosphorus, at ambient temperature and pressure. It is a layered material, with each layer forming a puckered surface due to $sp^3$ hybridization. In its bulk crystalline form \cite{Keyes53,Warschauer63,Jamieson63,morita86review,chang86}, BP is a semiconductor with a direct band gap of about $0.3\,$eV with measured Hall mobilities in $n$ and $p-$type samples approaching $10^5\,$cm$^2$/Vs. Recent rediscovery of BP \cite{Reich14com,Li14BP,Liu14BP,xia14bp,Koenig14,Castellanos14} in its multilayer  form revealed highly anisotropic electrical and optical properties. 

In this paper we examine the collective electronic excitations of BP, and its electrostatic screening behavior both along the in- and out-of-plane directions.
We calculate the dielectric function $\epsilon(\textbf{q},\omega)$, at finite frequency $\omega$ and wavevector $\textbf{q}$, for monolayer and multilayers of BP within the Random Phase Approximation (RPA), using an effective low-energy Hamiltonian \cite{rodin14}. The inherent anisotropy of screnning is studied, and the out-of-plane screening properties of multi-layer BP flakes are considered within a non-linear Thomas-Fermi model. The 2D plasmon modes, which are obtained from the zeros of the dielectric function, or the electron loss spectra, $\Im[-1/\epsilon(\textbf{q},\omega)]$, show a highly anisotropic plasmon dispersion, $\omega_{pl}(\textbf{q},\omega)$, and we studied its scaling behavior with doping. Lastly, we discuss the implications of our results for basic electrical and light scattering experiments.

\emph{Hamiltonian---} BP has an orthorhombic crystal structure consisting of puckered layers. The lattice constant in the out-of-plane direction is about $10.7\,$\AA, and the effective layer-to-layer distance is half of this value \cite{morita86review}. In monolayer BP, translational symmetry in the $z$ direction is broken, and its bandstructure has a direct energy gap at the $\Gamma$ point instead of the Z point in the bulk case. Based on $\mathbf{k}$$\cdot$$\mathbf{p}$ theory and symmetry arguments, the in-plane electron dispersion around the $\Gamma$ point can be described by the following low-energy Hamiltonian \cite{rodin14},
\begin{eqnarray}
{\cal H} = \left(
\begin{array}{cc}
E_c + \eta_c k_x^2 + \nu_c k_y^2 & \gamma k_x + \beta k_y^2\\
\gamma k_x + \beta k_y^2 & E_v - \eta_v k_x^2 - \nu_v k_y^2
\end{array} \right)
\label{hamil}
\end{eqnarray}
where $\eta_{c,v}$ and $\nu_{c,v}$ are related to the effective masses, while $\gamma$ and $\beta$ describe the effective couplings between the conduction and valence bands. $E_c$ and $E_v$ are the energies of the conduction and valence band edges. At present, the energy gap for monolayer BP has not been measured experimentally, but recent \emph{ab initio} calculation based on the GW method found an energy gap of $\sim 1.5-2\,$eV \cite{Tran14,rudenko14}.

Unlike other layered materials such as graphene and the transition metal dichalcogenides (TMDs), electrons in BP are energetically highly dispersive along the out-of-plane direction. Cyclotron resonance experiments on bulk BP \cite{Narita83cy} found an out-of-plane effective mass considerably smaller than that of TMDs \cite{Mattheiss73}. For multilayer BP, confinement in the out-of-plane $z$ direction leads to multiple subbands. The in-plane dispersion within each subband $j$ can be described by Eq.\,\eqref{hamil}, where $E_{c,v}$ are being replaced with $E_{c,v}^j$. More explicitly, $\delta E_{c}^j$ is given by $j^2\hbar^2\pi^2/2m_{cz}d^2+\delta_c(d)$, where $j$ labels the subband, $d$ is the thickness of the BP film, and $m_{cz}$ is the electron effective mass along $z$. Analogous expressions apply also for the hole case. The quantities $\delta_{c,v}(d)$ are chosen such that it reproduces the energy gap of the BP film \cite{Tran14}, of 2\,eV and 0.3\,eV in the monolayer and bulk limit respectively. In this work, we adopt an average of experimental \cite{Narita83cy} and theoretically \cite{Narita83cy,low14bpcond} predicted quantization mass i.e. $m_{cz}\approx 0.2\,m_0$ and $m_{vz}\approx 0.4\,m_0$.

The in-plane dispersion is mainly determined by the parameters $\eta_{c,v}$, $\nu_{c,v}$ and $\gamma$. These parameters are chosen such that they yield the known anisotropic effective masses. In the bulk BP limit, we have $m_{cx}=m_{vx}=0.08\,m_0$, $m_{cy}=0.7\,m_0$ and $m_{vy}=1.0\,m_0$ \cite{morita86review,Narita83cy}, and $m_{cx}=m_{vx}\approx 0.15\,m_0$ for monolayer BP \cite{low14bpcond}. Using this knowledge, we arrive at the following parameter set; $\eta_{c,v}=\hbar^2/0.4m_0$, $\nu_{c}=\hbar^2/1.4m_0$, $\nu_{v}=\hbar^2/2.0m_0$, and $\gamma=4a/\pi\,$eVm. The value of $\beta$ is taken to be $\approx 2a^2/\pi^2\,$eVm$^2$ \cite{low14bpcond}, where $a\approx 2.23$ \AA\, and $\pi/a$ is the width of the BZ in $x$ direction.


\begin{figure}[t]
\centering
\scalebox{0.7}[0.7]{\includegraphics*[viewport=150 270 465 470]{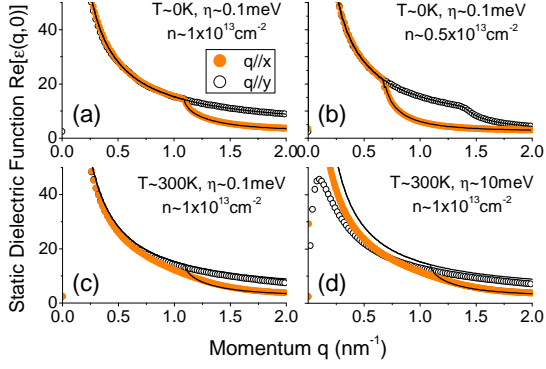}}
\caption{\textbf{Static screening.} \textbf{(a-d)} Dielectric function in the static limit, for different conditions as stated. Solid black lines are analytical expression for $\epsilon(q,0)$ at $T\rightarrow 0\,$K and $\eta\rightarrow 0\,$eV.
}
\label{fig1}
\end{figure}

\emph{Dielectric function---}
The dielectric function of an electron gas in the RPA can be written as,
\begin{eqnarray}
\epsilon(\textbf{q},\omega) = \kappa + v_c(q)\Pi(\textbf{q},\omega)
\end{eqnarray}
where $v_c(q)=e^2/2\epsilon_0 q$ is the 2D Coulomb interaction and $\kappa$ describes the effective dielectric constant of the medium, which for a common substrate, SiO$_2$, is $\sim$$2.5$. $\Pi(\textbf{q},\omega)$ is the 2D polarizability (i.e. the pair bubble diagram) given by,
\begin{eqnarray}
\nonumber
\Pi(\textbf{q},\omega)&=&-\frac{g_s}{(2\pi)^2}\sum_{ss'jj'}\int d\textbf{k} \frac{f_{sj\textbf{k}}-f_{s'j'\textbf{k}'}}{E_{sj\textbf{k}}-E_{s'j'\textbf{k}'}+\hbar\omega+i\eta}\\
&&\times|\left\langle \Phi_{sj\textbf{k}} \right.\left|\Phi_{s'j'\textbf{k}'}\right\rangle|^2
\label{pi}
\end{eqnarray}
where $\textbf{k}'=\textbf{k}+\textbf{q}$, $\{s,s'\}=\pm 1$ denote conduction/valence bands, while $\{j,j'\}$ are the subband indices and $g_s=2$ is the spin degeneracy. $E_{sj\textbf{k}}$ and $\Phi_{sj\textbf{k}}$ are the eigen-energies and eigen-functions after diagonalizing ${\cal H}$.
$f_{sj\textbf{k}}=\{\mbox{exp}[(E_{sj\textbf{k}}-\mu)/k_BT]+1\}^{-1}$ is the Fermi distribution function, where $\mu$ is the chemical potential. Finite damping can be modeled with the phenomenological broadening term $\eta$. Allowed optical transitions between these quantized subbands occur when $ss'=\pm 1$ (i.e. intra- and inter-band processes) and $j=j'$. Otherwise the matrix element $\left\langle ...\right\rangle$ in Eq.\,\eqref{pi} vanishes.


\emph{Screening---}
In the static limit, we generalize the well-known analytical form of the polarizability for 2D electron gas (2DEG) \cite{Lindhard54} to include anisotropy. Since $\omega=0$, we deal only with intraband processes. In the $T\rightarrow 0$ and $\eta\rightarrow 0$ limits, we have \cite{kittel04},
\begin{eqnarray}
\nonumber
\Pi(\textbf{q})&=&-\frac{g_s}{(2\pi)^2}\int_0^{\textbf{k}_F}d\textbf{k} \left[\frac{1}{E_{\textbf{k}+\textbf{q}}-E_{\textbf{k}}}-\frac{1}{E_{\textbf{k}}-E_{\textbf{k}-\textbf{q}}}\right]\\
&=& -\frac{g_s m_d}{\pi^2\hbar^2} \int_{0}^{p_F} dp \int_{0}^{2\pi}d\theta \frac{p}{s^2-4p^2\mbox{cos}^2\theta}
\end{eqnarray}
where we have made the transformation,
\begin{eqnarray}
\textbf{k}\rightarrow \frac{1}{\sqrt{m_d}}M^{\tfrac{1}{2}} \textbf{p} & \mbox{and} & \textbf{q}\rightarrow \frac{1}{\sqrt{m_d}}M^{\tfrac{1}{2}} \textbf{s}
\end{eqnarray}
where $M$ is the mass tensor with diagonal elements $m_x$ and $m_y$, $m_d$ is the 2D density-of-states mass given by $\sqrt{m_x m_y}$, and $p_F=\sqrt{2m_d\mu}/\hbar$. After some algebra, we arrive at
\begin{eqnarray}
\Pi(\textbf{q}) = g_{2D}\Re \left[1-\sqrt{1-\frac{8\mu/\hbar^2}{q_x^2/m_x+ q_y^2/m_y}}\right]
\label{staticscreen}
\end{eqnarray}
where $g_{2D}=m_d/\pi\hbar^2$ is the 2D density-of-states.
We make an interesting remark: for $q\leq 2|\textbf{k}_F\cdot \hat{\textbf{q}}|$, we see that $\Pi(\textbf{q})$ reduces to the familiar relation for the static polarization of a 2DEG, $\Pi(\textbf{q})=g_{2D}$. Long range potentials, such as those induced by charged impurities, involve momenta $q$ such that  $q\leq 2|\textbf{k}_F\cdot \hat{\textbf{q}}|$, so that screening will be isotropic, at least in the zero temperature and disorder limits.

Fig.\,\ref{fig1}(a) compares the static dielectric function obtained numerically with the analytical model in Eq.\,\eqref{staticscreen}, with excellent agreement in the limits of the model. $\Pi(\textbf{q})$ has a kink at $q= 2|\textbf{k}_F\cdot \hat{\textbf{q}}|$. Fig.\,\ref{fig1}(b) illustrates how the kink migrates with change in doping. With increasing temperature and disorder, the kink is smoothed out as illustrated in Fig.\,\ref{fig1}(c)-(d), showing obvious deviation from the analytical model. The otherwise isotropic screening at small momenta now becomes anisotropic.
On the other hand, dynamical screening, $\epsilon(\textbf{q},\omega)$, in BP exhibits strong directional dependence with $\textbf{q}$. Anisotropic dynamic screening might have important implications to carrier relaxation processes such as scattering with polar optical phonons. In Suppl. Info, we show the calculated real and imaginary part of $\epsilon(\textbf{q},\omega)$ at finite $\omega$. 

\begin{figure}[t]
\centering
\scalebox{0.7}[0.7]{\includegraphics*[viewport=150 230 465 640]{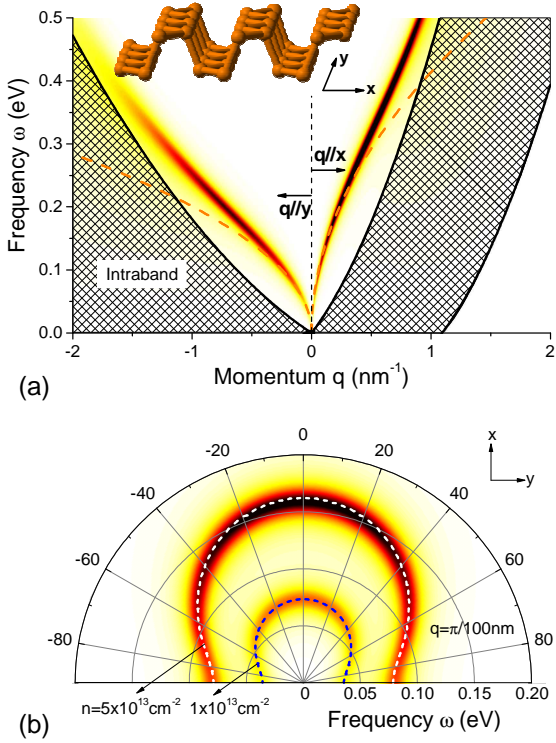}}
\caption{\textbf{Energy loss and plasmon dispersion.} \textbf{(a)} Loss function, $L(\textbf{q},\omega)=-\Im[1/\epsilon(\textbf{q},\omega)]$, calculated for monolayer BP for electron doping of $1\times 10^{13}\,$cm$^{-2}$, for $q$ along the two crystal axes $x$ (right) and $y$ (left). We assumed $T=300\,$K and $\eta=10\,$meV. Shaded regions are the Landau damping regions, defined by the single particle phase space $\hbar\omega_{SP}^{\pm}(\textbf{q})$ as described in text.
\textbf{(b)} Polar intensity plot of $L(\textbf{q},\omega)$ for electron doping of $1\times 10^{13}\,$cm$^{-2}$ and  $5\times 10^{13}\,$cm$^{-2}$ under same conditions as (a), for specified $q$. The radial and azimuth coordinates denote the frequency $\omega$ and the angular orientation of $q$. Dashed lines are the plasmon solutions from Eq.\,\ref{mweq1}.
}
\label{fig2}
\end{figure}

\emph{Plasmon dispersion---}
The zeros of the dynamical dielectric function $\epsilon(\textbf{q},\omega)$ yield the excitation spectrum of the plasmon modes of the electron gas. The loss function, defined as $L(\textbf{q},\omega)=-\Im[1/\epsilon(\textbf{q},\omega)]$, quantifies the spectral weight of the plasmon mode, which presents itself as a delta peak in the limit of zero damping. Experimentally, $L(\textbf{q},\omega)$ can be quantified with EELS. In the long wavelength limit, i.e. $q\ll k_F$, these modes are well-described by classical Maxwell theory. We consider a BP film sandwiched between two dielectric media $\epsilon_1$ and $\epsilon_2$. The bound modes, i.e. plasmons, are characterized by an in-plane wavevector $\textbf{q}$ pointing at an angle $\theta$ with respect to $x$. The dispersion relation for the bound mode can be obtained from the solution to the following equation,
\begin{eqnarray}
(\bar{Y}_{s}+M_{ss})(\bar{Y}_{p}+M_{pp})+M_{ps}M_{sp}=0
\label{mweq1}
\end{eqnarray}
where $\bar{Y}_{\beta}=Y_{\beta}^1+Y_{\beta}^2$ ($\beta=s,p$) is the total admittance, with $Y^i_s=Y_0 (k_{zi}/k_0)$ and $Y^i_p=Y_0 \epsilon_i (k_0/k_{zi})$,
and $k_{zi}^2=k_0^2\epsilon_i-q^2$, $k_0=\omega/c$. $c$ and $Y_0=\sqrt{\epsilon_0/\mu_0}$ are the speed of light and admittance of free space, respectively. The matrix elements of $M$ are expressed in terms of $\sigma_{jj}$, the diagonal components of the 2D BP conductivity tensor,
\begin{eqnarray}
\nonumber
M_{ss}&=&\sigma_{xx}\mbox{sin}^2\theta+\sigma_{yy}\mbox{cos}^2\theta\\
\nonumber
M_{sp}&=&M_{ps}=(\sigma_{yy}-\sigma_{xx})\mbox{sin}\theta\mbox{cos}\theta\\
M_{ss}&=&\sigma_{xx}\mbox{cos}^2\theta+\sigma_{yy}\mbox{sin}^2\theta
\end{eqnarray}
In the limits $\theta=0,\pi$ and $\sigma_{xx}=\sigma_{yy}$, Eq.\,\ref{mweq1} reduces to
\begin{eqnarray}
\bar{Y}_{p}+M_{pp}=0
\end{eqnarray}
In the non-retarded regime, i.e. $q\gg k_0$, hence $k_{zi}\approx iq$, we obtain the `quasi-static' approximation,
\begin{eqnarray}
-\frac{\sigma_{xx}\mbox{cos}^2\theta+\sigma_{yy}\mbox{sin}^2\theta}{\epsilon_0 \omega}=\frac{\epsilon_1}{k_{z1}}+\frac{\epsilon_2}{k_{z2}}\approx \frac{2\kappa}{iq}
\label{maxwellsol}
\end{eqnarray}
where $\kappa=(\epsilon_1+\epsilon_2)/2$. For frequencies up to the mid-infrared, the conductivity can be approximated by the Drude model,
\begin{eqnarray}
\sigma_{jj}(\omega)=\frac{i {\cal D}_{j}}{\pi(\omega+i\eta/\hbar)} &\mbox{   ,   }& {\cal D}_{j}=\pi e^2 \sum_{i}\frac{n_i}{m_{j}^i}
\label{conddr}
\end{eqnarray}
where ${\cal D}_{j}$ is the Drude weight and $i$ denotes the subbands. Within the model Hamiltonian, the in-plane electron effective masses in vicinity to the $\Gamma$ point are given by the following expressions \cite{rodin14},
\begin{eqnarray}
m_{cx}^i=\frac{\hbar^2}{2\gamma^2/\Delta^i+\eta_{c}} &\mbox{   ,   }& m_{cy}=\frac{\hbar^2}{2\nu_c}
\end{eqnarray}
where $\Delta^i$ is the subband energy gap. Similar expressions apply for the hole case. Note that in graphene, ${\cal D}=\mu e^2/\hbar^2$ instead. With Eq.\,\eqref{maxwellsol} and \eqref{conddr}, we have the classical plasmon dispersion along the $j=x,y$ directions, which is $\omega_{pl,j}(\textbf{q})=\sqrt{({\cal D}_j /2\pi\epsilon_0\kappa)q}$.

\begin{figure}[t]
\centering
\scalebox{0.7}[0.7]{\includegraphics*[viewport=150 385 465 600]{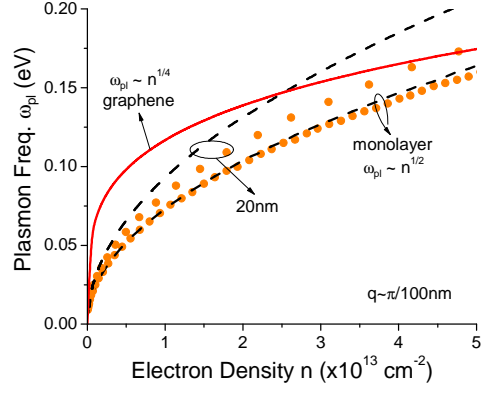}}
\caption{\textbf{Plasmon scaling with carrier concentration.} Plasmon energies, $\omega_{pl}$, as function of density $n$, calculated for the monolayer and for a $20\,$nm BP thick film at a specified $q$ along $x$. Graphene plasmons are shown for comparison. Dashed lines are the long-wavelength estimates using Eq.\,\eqref{maxwellsol} and \eqref{conddr}. }
\label{fig3}
\end{figure}


\begin{figure}[t]
\centering
\scalebox{0.7}[0.7]{\includegraphics*[viewport=150 175 465 570]{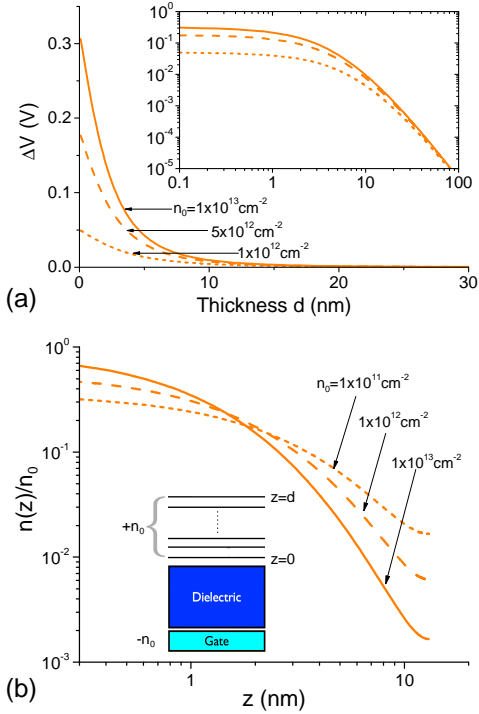}}
\caption{\textbf{Interlayer screening. (a)} Potential difference $\Delta V(d)$ as a function of the thickness for an electron doped sample, obtained from the Thomas-Fermi model. We show the results for different carrier concentrations $n_0$. The insets show the same results in a logarithmic scale. \textbf{(b)} Normalized surface-charge distribution $n(z)$ induced in a $13\,$nm thick sample for different gate carrier densities $n_0$ as stated, and inset shows a sketch of the model. }
\label{fig4}
\end{figure}

Fig.\,\ref{fig2}(a) plots the RPA loss function $L(\textbf{q},\omega)$ for momentum along the two main crystallographic directions for monolayer BP, with an electron doping of $10^{13}\,$cm$^{-2}$. The plasmon disperses differently due to their mass anisotropy, where the smaller mass along $x$ leads to higher resonance frequency. Classical plasmon dispersion agrees well with the RPA result in the long wavelength limit. Due to the energy gap of $2\,$eV for monolayer BP, Landau damping occurs preferentially via intraband processes. This occurs when plasmon enters the SP phase space, whose boundaries are given by, $\hbar\omega_{SP}^{\pm}(\textbf{q})=E(\pm \textbf{k}_{F}+\textbf{q})-E(\textbf{k}_{F})$.
Our calculation suggests that the plasmon along the $y$ direction is damped at mid-infrared frequencies, while the plasmon along $x$ persists up to the near infrared.

The results reported here can be tested by EELS. In addition, plasmon modes in layered materials \cite{grigorenko12,lowreviewacs,Stauber14} can also be probed by Fourier transform infrared (FTIR) light scattering experiments of nanostructures \cite{Yan13damping,ju11} or with infrared nano-microscopy techniques \cite{fei12gate,chen12op}. For example, nanostructures exhibit prominent resonances in their extinction spectra due to localized plasmons with odd multiple of the momentum $q=\pi/W$ where $W$ can be the width of nanoribbons, or the diameter of nano-disks. Fig.\,\ref{fig2}(b) shows $L(\textbf{q},\omega)$ for different angular orientation of $\textbf{q}$ for a momentum corresponding to nanostructures of $100\,$nm in size. Dashed lines are solutions of Eq.\,\ref{mweq1}. The results suggest polarization sensitive mid-infrared plasmonic resonances in the absorption spectra in BP nanostructures.

Fig.\,\ref{fig3} studies the scaling of plasmon frequency (along $x$) with carrier concentration $n$. For monolayer BP, we obtain the expected scaling relation of $\omega_{pl}\propto n^{1/2}$, as in conventional 2DEGs. However, for thicker samples, we found that $\omega_{pl}\propto n^{\beta}$, with $\beta<\tfrac{1}{2}$ instead. This deviation is due to the strong non-parabolicity caused by interband coupling, particularly when the energy gap of the BP film is $\ll \gamma^2/\eta_{c,v}$. Hence, non-parabolicity effects are more prominent for thicker films. We also note the general trend of increasing Drude weight with film's thickness due to the decreasing effective masses (see Suppl. Info.).

\emph{Screening and charge distribution in multilayers---}
We complete our study by considering the charge distribution and the electrostatic screening in few-layer BP sheets. For this aim we use a  non-linear TF theory, which has been shown to properly account for the screening properties of graphite \cite{Pietronero78,Datta08scre,Kuroda11scr} and MoS$_2$ \cite{Castellanos13}. We start by considering a given charge transfer between the substrate and the BP flakes, whose origin can be due to charge impurities in the substrate or to the action of a gate voltage. This charge transfer leads to a net surface charge density $en_0$ while a layer below the substrate acquires a charge of $-en_0$, see inset of Fig.\,\ref{fig4}(b). For a BP sample of thickness $d$, the electrostatic potential $V(z)$ and the carrier distribution $n(z)$ as a function of the distance from the substrate $z$ can be obtained from the energetic balance between kinetic and interlayer capacitance terms, which leads to the non-linear differential equation \cite{Castellanos13}
\begin{equation}
\frac{d^2f(z)}{dz^2}=\frac{5}{2}\beta_{\perp}f(z)^{3/2}
\end{equation}
where $f(z)=[en(z)]^{2/3}$ and we have defined $\beta_{\perp}=(4e^2/5\epsilon_0\kappa)(g_sd_0m_{d}\sqrt{m_{z}}/6\pi^2\hbar^3)^{2/3}$, where $d_0\approx 1.07\,$nm and $\kappa\approx 8.3$ are the interlayer separation and dielectric constant, respectively \cite{morita86review}. Using the boundary conditions $f'(0)=\frac{5}{2}\beta_{\perp}en_0$ and $f'(d)=0$, one can obtain the charge density from the solution of the integral equation
\begin{equation}
\int_{f(0)}^{f(z)}\frac{df}{\sqrt{f^{5/2}-f^{5/2}(d)}}=\sqrt{\frac{2\beta_{\perp}}{d_0}}z.
\end{equation}
On the other hand, the potential difference across a BP sample of thickness $d$ can be shown to be given by \cite{Castellanos13}:
\begin{equation}
\Delta V(d)=\frac{2e^2}{5\epsilon_0\kappa\beta_{\perp}^{3/5}}\left(\frac{25d_0 e^2n_0^2}{8}\right)^{2/5}\frac{1-r_d}{\left(1-r_d^{5/2}\right)^{2/5}}
\end{equation}
where we have defined the dimensionless parameter $r_d=n^{2/3}(d)/n^{2/3}(0)$.

The potential difference obtained from the above model is shown, for different carrier concentrations, in Fig.\,\ref{fig4}(a) for a n-doped sample (see Suppl. Info. for results also on p-doped samples). The screening of charged impurities or the gate potential increases as the thickness of the BP layer grows. The dependence of $\Delta V(d)$ on $d$ suggests an intermediate screening behavior between the strong coupling limit of graphene, where the carriers concentrate close to the interface \cite{Datta08scre}, and the weak coupling regime with reduced screening properties that dominates the screening of MoS$_2$ \cite{Castellanos13}. \textcolor{black}{Our results suggest that the gate will have negligible effect $10\,$nm into the bulk of BP, consistent with recent experiments on multilayers BP transistors \cite{xia14bp}. }
We have also calculated $n(z)$ for a sample with a given thickness $d$ but different charge carrier concentrations $n_0$ as shown in Fig.\,\ref{fig4}(b). We observe a strong dependence of the screening strength on $n_0$, such that stronger screening is achieved for larger $n_0$. Interestingly, from those results one could infer a screening length of the order of the inter-layer spacing for $\sigma_0=10^{13}{\rm cm}^{-2}$, whereas for lower concentrations, like $10^{11}{\rm cm}^{-2}$, the screening length is one order of magnitude larger.

\emph{Conclusions---} In conclusion, we have studied the screening properties of BP using a combination of RPA for the dynamic and static in-plane screening, as well as for the dispersion of the collective (plasmon) excitations, and a non-linear TF theory for the inter-layer screening. Whereas we find a relatively isotropic static screening, the band non-parabolicity leads to highly anisotropic plasmons. Most saliently, we find that for multilayer samples, the plasmon resonance scales with doping as $n^{\beta}$, where $\beta<\tfrac{1}{2}$. Furthermore, the modes dispersing along one of the crystallographic directions are long lived, being Landau damped (i.e. decaying into intra-band electron-hole pairs) only for high frequencies, near the infrared. Finally, we find that the charge distribution along the layers and the strength of the electric field screening in BP flakes seem to be between the strong coupling regime characteristic of graphene, and the weak coupling regime of the TMD semiconductors, such as MoS$_2$.

\emph{Acknowledgements---} FG and RR acknowledge support from the Spanish Ministry of Economy (MINECO) through Grant No. FIS2011-23713, the European Research Council Advanced
Grant (contract 290846), and the European Commission under the Graphene Flagship, contract CNECT-ICT-604391. R.R. acknowledges financial support from the Juan de la Cierva Program. 


\end{document}